\documentclass[aps,prl,twocolumn,showpacs]{revtex4}
\usepackage{amssymb}
\usepackage{graphicx}
\usepackage{epsfig}

\usepackage{dcolumn}
\usepackage{bm}

\newcommand{\be}{\begin{equation}}
\newcommand{\ee}{\end{equation}}
\newcommand{\bea}{\begin{eqnarray}}
\newcommand{\eea}{\end{eqnarray}}
\newcommand{\<}{\left\langle}
\renewcommand{\>}{\right\rangle}

\newcommand{\p}{\partial}
\newcommand{\nn}{\nonumber }

\usepackage{color}

\definecolor{db}{cmyk}{1,0.4,0,0.4}
\definecolor{dr}{cmyk}{0,0.9,0.7,0.4}
\definecolor{dg}{cmyk}{1,0,1,0.2}

\begin{document}

\date{\today}
\title{Collective modes in quantum electron glasses and electron-assisted hopping}
\author{M.\ M\"uller}
\affiliation{Department of Physics, Harvard University, Cambridge MA 02138}
\author{L. B. Ioffe}
\affiliation{Department of Physics, Rutgers University, Piscataway, New Jersey 08854 }

\begin{abstract}
We study electronic transport in Anderson insulators with strong Coulomb interactions in dimensions $d\geq 2$. Close to the metal insulator transition where the single particle localization length is
much larger than interparticle-distance,
the interactions lead to a strongly correlated quantum
glass phase. Even though single particle excitations are localized and the system is insulating, there are collective electronic modes which remain delocalized down to parametrically small energies.
These collective excitations serve as a continuous bath which can provide the activation energy for variable range hopping transport. This circumvents the energy conservation problem arising when only discrete particle-hole excitations are present. In contrast to the weak and material-dependent phonon-assisted hopping mechanism, the activation by an electronic bath leads to a nearly universal prefactor $e^2/h$ of the Efros-Shklovskii conductance, as is observed in many recent experiments.
\end{abstract}

\maketitle

Close to the metal-insulator transition (MIT), Coulomb interactions turn Anderson insulators into strongly correlated and frustrated quantum systems. Their transport properties have been a puzzle for a long time and are still lively debated. On the experimental side there is a longstanding discrepancy between the observed, almost universal pre-exponential factor of the hopping conductance~\cite{HoppingExp} (of order $e^2/h$ in two dimensions), and the much smaller and material-dependent value predicted by the standard theory based on the activation of electrons by phonons~\cite{MottBook}.
The universality of the prefactor lead to speculations about a purely electronic activation mechanism~\cite{Aleiner94,Berkovits99}, but a consistent microscopic theory remains elusive.
In fact, the situation became even less clear recently, when several authors showed that purely electron-assisted transport is impossible because of localization in Fock space, if interactions are
weak and sufficiently short ranged~\cite{Fleishman78,Gornyi05}.
The paper~\cite{Burin06} claims, however, that this dichotomy is resolved by the long range nature of Coulomb interactions, rendering Refs.~\cite{Fleishman78,Gornyi05} inapplicable to physical systems in $d\geq 2$, but it leaves the question of the prefactor open.
So does the theory of electron-assisted variable range hopping constructed for disordered Luttinger liquids in $d=1$~\cite{Nattermann03}, which, in addition, relies on the presence of a thermal phonon bath and thus is in fact not purely electronic in nature.

However, even without invoking long-range interactions, many body localization is unlikely to occur if the interactions are strong enough, as suggested by exact diagonalization studies in one- and two-dimensional models~\cite{Berkovits99,HuseOganesyan07}.
This is indeed what should be expected as the MIT is approached and single electron wavefunctions start to overlap: The strength of Coulomb interactions decreases less rapidly than the level spacing in a localization volume. This drives the electronic system to a strong coupling regime, where perturbation theory breaks down and a collective, quantum glassy state emerges. In this paper, we analyze this regime and show that, even though the system remains insulating, it possesses a continuum of delocalized collective electronic modes whose coupling to hopping electrons entails a nearly universal pre-exponential factor in the hopping conductivity, of the order observed in experiments.

We consider interacting electrons in $d\geq 2$, with disorder such that the single-particle localization length $\xi$ is much bigger than the average interparticle spacing, i.e.,
$n\xi^d \gg 1$, where $n$ is the electron density. As a basis for the interacting problem we choose the localized eigenfunctions $\psi_i(r)$ of the selfconsistent disorder potential, which accounts for the screening due to the "frozen" electron density common to all metastable states of the system. 
 In this basis the Hamiltonian reads
\bea
\label{Hamiltonian}
H&=&\frac{1}{2}\sum_{i\neq j}  n_{i} J_{ij} n_j +\sum_{i}(\epsilon_{i}-\mu) n_i\\
&&+\sum_{i\neq k,j} t_{i;j}^{k} c^\dagger_{k} c_{i} \delta n_j
+\frac{1}{2}\sum_{i,j\neq k,l} V_{ij}^{kl} c^\dagger_{k} c_{i}c^\dagger_{l}c_j ,\nn
\eea%
where $\mu$ is the chemical potential and $\delta n_i\equiv n_i-\<\<n_i\>\>$ is the deviation of the occupation number $n_i$ from the "frozen" background occupation $\<\<n_i\>\>$. The $\epsilon_i$ denote the energies of the non-interacting problem characterized by a bare density of states $\nu_0$. The typical level spacing in a localization volume is thus $\delta_\xi=1/(\xi^d\nu_0)$. $J_{ij}$ and $t_{i;j}^k, V_{ij}^{kl}$ are the diagonal and off-diagonal matrix elements of the Coulomb interactions, respectively. Due to the random charge density pattern $|\psi_i^2(r)|$ associated with the level $i$, the Coulomb interaction between two states within the same localization volume is a random quantity of order $|J_{ij}|\sim J\equiv e^2/\kappa\xi$, after subtraction of an irrelevant average repulsion. Here, $\kappa$ is the dielectric constant of the host. The strength of Coulomb interactions is measured by the ratio
\bea
\label{z}
z\equiv J/\delta_\xi =\nu_0 \xi^{d-1} e^2/\kappa,
\eea
which diverges as the  MIT is approached ($\xi\to \infty$). Once the wavefunctions start overlapping, $z$ rapidly becomes large, $z\gg 1$, leading to a highly frustrated glassy problem. 
This situation is opposite to the limit $z\ll 1$ analyzed in Refs.~\cite{Gornyi05}, which allowed for a perturbative treatment starting from the non-interacting ground state and lead to the prediction of many body localization.

The first line in (\ref{Hamiltonian}) defines a classical problem equivalent to an Ising spin glass with random fields.  Note that $z^2\gg 1$ plays the role of a large effective coordination number~\cite{LebanonMueller05,DotsenkoFeigelmanIoffe}, which allows for controlled approximations.
The classical Hamiltonian possesses many metastable states, i.e., low energy configurations $n_i$ which are stable with respect to the rearrangement of few electrons. The Hartree interactions with other electrons strongly modify the distribution of the effective energies of the levels, $\varepsilon_i=\epsilon_i-\mu +\sum_j J_{ij}n_j$, creating a linear pseudogap similar to the Efros-Shklovskii Coulomb gap~\cite{efrosshklovskii7576,TAP} which extends up to the large energy scale $E_C\sim zJ$~\cite{ButkoAdams00}. The Coulomb interactions reshuffle the single particle energies, and in a given localization volume the average level spacing at the Fermi level is renormalized to $J\gg \delta_\xi$. At temperatures of order $J$ one thus expects the onset of activated hopping conductivity.

In the presence of quantum fluctuations due to the off-diagonal terms of the Coulomb interactions, stability
 arguments show that the pseudogap in the quasiparticle energy distribution remains intact.
Further, according to standard estimates \cite{LeeRamakrishnan85} the matrix elements $t_{i;j}^k$ and $V_{ij}^{kl}$ decrease  with bare energy transfer on average like
\bea
\label{tikj2}
|t_{i;j}^k| &\sim & J \Phi\left( (|\epsilon_{i}-\epsilon_k|)/\delta\right),\\
\label{Vijkl}
|V_{ij}^{kl}| &\sim &  J \Phi\left( (|\epsilon_{i}-\epsilon_k|)/\delta\right)
\Phi\left( (|\epsilon_{j}-\epsilon_l|)/\delta\right),
\eea
with $\Phi(x)\sim x^{-1/2}$ for $x<(\xi/\ell)^d$, where $\ell\ll \xi$ is the mean free path.

The large effective coordination number $z^2$ and the slow decay of matrix elements ensure that a given level interacts with many neighbors. This allows us to neglect fermion loops and adopt a bosonic pseudo-spin $1/2$ description $n_i\to s_i^z+1/2$, $(c_i^\dagger,c_i)\to \sigma_i^{+,-}$, and to look for a self-consistent mean-field description of quantum fluctuations with the effective action (similarly to Refs.~\cite{pastor99})
\bea
S_{\rm eff}&=&\int_0^\beta d\tau\,\left[ \frac{1}{2}\sum_{i,j} \sigma^z_i(
\tau) J_{ij} \sigma^z_j(\tau)  +\sum_{i}(\epsilon_{i}-\mu) \sigma^z_i(\tau)\right]\nn\\
&& +\sum_i \int_0^\beta d\tau' \int_0^\beta d\tau\, \sigma^+_{i}(\tau') G_i(\tau'-\tau)\sigma^-_{i}(\tau).
\label{SpinHamiltonian2}
\eea%
The kernels $G_i(\tau)$ are due to the virtual hopping of electrons to other sites and back. Thus, they describe inertial, non-dissipative dynamics, which does not lead to significant level broadening, as we will verify below.
In order to gain insight into the collective modes of this system we proceed similarly as in the TAP approach to classical spin glasses~\cite{TAP,DotsenkoFeigelmanIoffe}, which is a good approximation for $z\gg 1$. The effective field acting on spin $i$ is
\bea
\label{TAP}
h_i 
=\epsilon_i-\mu +\sum_{j\neq i}J_{ij}m_j-m_i\sum_{j\neq i}J^2_{ij}\chi_j,
\eea
where the "magnetizations" $m_i=\<\sigma^z_i\>$ are averages restricted to a given metastable state.
The last term in (\ref{TAP}) describes the Onsager reaction of the environment of neighboring spins, where $\chi_j$ is the susceptibility of spin $j$.
An exact treatment would require to formulate self-consistent {\em dynamic} TAP equations
involving the dynamic susceptibilities $\chi_j(\tau-\tau')$~\cite{BiroliCugliandolo01}. However, the short-time dynamics describing collective modes can be captured by the much simpler static approximation~\cite{TAPcomment} (\ref{TAP}), which can be derived as saddle point equations $d{\cal F}(\{m_j\})/dm_i=0$ of an approximate static energy functional following Ref.~\cite{Plefka82}.

The equations (\ref{TAP}) are closed when supplemented by a relation  determining the magnetization $m_i$ in the presence of a field $h_i$. This can be achieved
by realizing
that quantum fluctuations act like effective transverse fields of order $t_{\rm eff}\sim E_C$ for the spins. We have estimated $t_{\rm eff}$ by calculating the magnetizations of sites in strong local fields
with a perturbative calculation based on the Hamiltonian (\ref{Hamiltonian}), and comparing them with the corresponding expression resulting from an effective transverse field. The reduction in magnetization is
due to hybridizations with other levels, predominantly mediated by the electron-electron scattering term $V^{kl}_{ij}$.
Since variations of the fluctuation strength 
are inessential for the physics we approximate the effective transverse field for simplicity by a constant magnitude $t_{\rm eff}\sim E_C$. This provides us with the sought relation
$h_i= 2 t_{\rm eff} m_i/[1/4-m_i^2]^{1/2}$ between field and magnetization, and 
$\chi_i=\frac{d m_i}{dh_i}=\frac{2}{t_{\rm eff}}\left[1/4-m^2_i\right]^{3/2}$ 
for the susceptibility.

The large magnitude of quantum fluctuations is due to the marginal decay of (\ref{tikj2},\ref{Vijkl}). This ensures a large number ${\cal N}_{\rm act}\approx t_{\rm eff}^2/\delta_\xi^2 \sim z^2$ of electrons per localization volume which participate significantly in the quantum dynamics of the glass, justifying our mean-field approximations.

In order to assess our approximate TAP analysis, we have tested it on the case of a fully connected spin model with random Gaussian couplings $J_{ij}$ ($\overline{J_{ij}^2}=J^2/N$) in a transverse field $t$, but without random fields ($\epsilon_i=0$). One finds that the instability of the paramagnetic TAP solution, $m_i=0$, occurs precisely at the same critical transverse field $t_c=2\sqrt{N}J$ as predicted by the static approximation of the replica approach~\cite{BrayMoore80}, which suggests that the two approximations are equivalent. Both underestimate the strength of fluctuations by approximating the Onsager response by the full adiabatic susceptibility of spins instead of their smaller dynamic response.

Solutions $\{m_i\}$ of Eqs.~(\ref{TAP}) represent metastable states of the quantum glass. Collective excitations around these local minima are
governed by their potential energy environment which is captured by the Hessian
\bea
\label{Hessian}
{\cal H}_{ij}&=& \frac{\p^2 {\cal F}(\{m_i\})}{\p m_i\p m_j}=\frac{\p h_i}{\p m_j}
\equiv -J_{ij}+L_i\delta_{ij}\,,
\eea
where we have defined the 'locator' $L_i=\chi_i^{-1}+\sum_{k}J^2_{ik}\chi_k$.
Similarly as in classical glasses~\cite{BrayMoore79} and in mean field quantum glasses~\cite{MillerHuse93,BiroliCugliandolo01}, one expects that for $J\ll t_{\rm eff}$,
the spectrum of ${\cal H}$ is gapless, with a density of eigenvalues
\bea
\label{rholambda}
\rho_{\cal H}(\lambda)=\frac{\sqrt{\lambda}}{\pi}
\left[\sum_j {\left(J_{ij}^2\chi_j\right)^3}\right]^{-1/2}
\sim \frac{\sqrt{\lambda t_{\rm eff}}}{J^2}\,,
\eea
for $\lambda\lesssim t_{\rm eff}$.
The gaplessness within the glass phase reflects the marginal stability of the quantum glass which self-organizes into states $\{m_i\}$
characterized by a critical Onsager response, $\sum_j J_{ij}^2\chi_j^2=1$, and an abundance of collective low energy modes.

For the purpose of understanding the low energy spectrum, the Hessian (\ref{Hessian}) can be restricted to the "quantum active" spins with $|h_i|< t_{\rm eff}$, recalling that any spin $i$ interacts substantially via $J_{ij}$ with ${\cal N}_{\rm act}\sim z^2$ active spins. The eigenvectors of random matrices with this structure have been analyzed in \cite{DotsenkoFeigelmanIoffe}. One finds that there is a parametrically small mobility edge $\lambda_c\sim J z^{-5/3}$ in the spectrum, above which all modes are delocalized.

The quantum dynamics of these collective excitations is non-dissipative at low energies, as we will check below. We thus assume them to behave like harmonic oscillators with restoring force $\lambda$, and a mass $M\sim t^{-1}_{\rm eff}$ independent of the mode (in the regime of interest $\lambda<t_{\rm eff}$), as will be justified later. The collective modes are thus characterized by a frequency $\omega_\lambda= (\lambda t_{\rm eff})^{1/2}$ and mean square displacement
$\<x^2\>_\omega \sim t_{\rm eff}/|\omega|$. From the density of restoring forces (\ref{rholambda}) we then easily obtain 
the spectral function of collective modes (for $\omega\ll t_{\rm eff}$)
\bea
\label{spectralfunction}
\chi''(\omega) =  \<x^2\>_{\omega} \int d\lambda \,\rho(\lambda)\,\delta(\omega-\omega(\lambda)) \sim \frac{|\omega|}{J^2},
\eea
which, remarkably, turns out to be independent of $t_{\rm eff}$.
Eq.~(\ref{spectralfunction}) is equivalent to a local dynamic density correlator with long time asymptotics $\xi^{2d}\<\rho(r,\tau)\rho(r,0)\> \sim 1/(J\tau)^2$. We have adjusted the oscillator mass $M$ in such a way that this correlator scales like ${\cal N}_{\rm act}$ for $\tau\sim M$, reflecting that all active spins take part in the dynamics at these short time scales.
The power law decay $\tau^{-2}$ of density correlators and linear spectral functions $\chi''(\omega)\sim |\omega|$ are well-known features of quantum glasses with undamped dynamics at and below the glass transition~\cite{MillerHuse93}.
Here we obtain this result with the physically intuitive interpretation of collective modes as underdamped harmonic oscillators governed by the semicircular distribution (\ref{rholambda}). We have verified that neither particle-hole excitations nor the coupling to other collective modes provide substantial friction, so that the description in terms of underdamped oscillators is indeed correct.

We emphasize that the collective modes with energies above $\omega_{\rm loc}\equiv \omega_{\lambda_c}\sim Jz^{-1/3}$ are delocalized, and hence form a bath with continuous spectrum. In linear approximation low energy modes with $\omega<\omega_{\rm loc}$ are localized, but it is possible that cubic couplings among them induce delocalization which might lower the edge $\omega_{\rm c}$ where the spectrum of collective modes turns discrete. Even though it seems unlikely, we cannot rule out that $\omega_{\rm c}=0$.

The collective modes are globally neutral charge density oscillations which cannot contribute directly to the electrical conductivity. However, they can activate single particle transitions close to the Fermi level, leading to {\em electron}-assisted hopping.
Note that the continuity of the spectrum is crucial to ensure energy conservation in an elementary hopping process. In contrast, particle-hole excitations are discrete and thus ineffective as a bath for inelastic hopping processes~\cite{Fleishman78,Gornyi05}.

It is not immediately obvious that in the presence of quantum fluctuations and collective low energy degrees of freedom the system remains
an insulator. However, we have checked that single particle-excitations remain localized.
Further, at low temperature the level broadening is smaller than the level spacing at the Fermi level in a localization volume. Indeed, decay rates of low-lying states are very small since they involve transitions with strongly suppressed matrix elements. On the other hand, pure dephasing due to the coupling of single particle levels to the collective bath leads to a level broadening of the order of $1/T_2\sim T$. This is negligible at temperatures $T\ll J$ where
 Efros-Shklovskii type variable range hopping sets in~\cite{efrosshklovskii7576,classicalCoulombgap}. We also note that the large parameter $z$ allows us to neglect the small overlap a single particle excitation has with a collective mode.

Following standard arguments of hopping theory~\cite{ESbook}, and replacing phonons by the electron bath with spectral function (\ref{spectralfunction}), one obtains a percolation problem of Miller-Abrahams resistors with elementary conductances
\bea
\label{gij}
g_{ij} &=& \tilde g(\Delta \varepsilon_{ij}) e^{-\frac{2|r_i-r_j|}{\xi}-\frac{\Delta \varepsilon_{ij}}{T}} n_F(\varepsilon_i)[1-n_F(\varepsilon_j)],
\eea
where $n_F$ is the Fermi distribution and $\Delta \varepsilon_{ij}=\varepsilon_j-\varepsilon_i-J_{ij}$ is the energy to transfer an electron from site $i$ to $j$.
For electron-assisted activation we find
$\tilde{g}(\Delta \varepsilon)  \sim  \frac{e^2}{h} \frac{\Delta \varepsilon}{T} \frac{J^2}{\Delta \varepsilon^2}$,
which depends only on the strength of Coulomb interactions, $J$. This is in marked contrast to phonon-assisted hopping which leads to a much smaller prefactor $\tilde g$ proportional to the material-dependent electron-phonon coupling constant.
The bulk conductivity follows from percolation arguments for the network of elementary resistors with conductances (\ref{gij})~\cite{ESbook}. For $d$-dimensional systems this yields the prediction
\bea
\label{sigma}
\sigma^{(d)} &\approx &\tilde{g}(E_{\rm hop})\xi^{2-d}\left(\xi/R_{\rm hop}\right)^{1+\nu_d}e^{-\left(J/T\right)^{1/2}}\nn\\
&\sim&
(e^2/h)\,\xi^{2-d}\left(J/T\right)^{\mu_d}e^{-\left(J/T\right)^{1/2}},
\eea
where the energy and distance of typical bottleneck hops are $E_{\rm hop}/T\sim R_{\rm hop}/\xi\sim (J/T)^{1/2}$.
With the percolation exponents $\nu_d \approx 1.34\, (0.9)$  for $d=2\,(3)$, respectively, one finds a weak $T$ dependence of the prefactor, with $\mu_d=3/2-(1+\nu_d)/2 =0.33\, (0.55)$.

Note that for $d=2$, our theory predicts
a nearly universal prefactor of order $e^2/h$ (as obtained by extrapolating conductivity plots to $J/T \rightarrow 0$), while for $d=3$ it is of the order of the minimal metallic conductivity. Such behavior has been reported in many recent experiments~\cite{HoppingExp}, 
the prefactor being consistent with small or vanishing $\mu_d\geq 0$.
Note however, that the latter is sensitive to assumptions about the matrix elements involved in elementary hopping processes, and is therefore less robust than the near universality of the prefactor.




We emphasize that the above derivation is valid only for temperatures such that the activation energy of bottleneck resistors exceeds the "mobility edge" in the spectrum of collective modes, $E_{\rm hop}(T)>\omega_{c}$.
At lower temperatures the conductivity becomes simply activated of the form $\sigma\propto  \exp(-\omega_{c}/T)$, but still with a prefactor of electronic origin. At even lower temperature, the phonon bath will take over and restore variable range hopping, however with a significantly smaller prefactor. Apart from the clear signature in the conductivity, the presence of delocalized collective electronic modes is expected to show experimentally
in their contribution to thermal conductivity~\cite{Burin89}, which should decrease substantially as the temperature drops below $\omega_{c}$.

In the derivation of the properties of the quantum glass we have tacitly assumed that the quantum fluctuations, as measured by $t_{\rm eff}$, are smaller than the critical value $t_c\sim E_C$ which would melt the glass~\cite{pastor99} and lead to a metallic state with delocalized quasiparticle excitations.
While it is difficult to imagine that this could result from repulsive Coulomb interactions, we cannot directly rule out this possibility at very large $z$.
However, we can infer the existence of a stable quantum glassy regime with overlapping electron wavefunctions ($z>1$), from the established existence of a glass phase~\cite{Cbglass} in the nearly classical regime of well localized electrons in moderate disorder where $z\approx 1$.
This conjecture is borne out by experimental observation of
 quantum glassy behavior in amorphous insulators with high carrier density~\cite{QuantumEG1}. Those systems can be taken into a regime of large $z$ where glassy effects in fact become stronger~\cite{QuantumEG2}. The puzzling density and temperature dependence of glassy relaxation reported in~\cite{QuantumEG2} might be related to the emergence of low energy excitations discussed in this paper.

We thank M. Feigelman, M. Gershenson, B. Halperin, D. Huse, D. Khmelnitskii,  and Z. Ovadyahu for many interesting and helpful discussions. This work was supported by SNF grant PA002-113151, ARO grant W911NF-06-1-0208 and NSF grant ECS 0608842.


\begin{thebibliography}{99}
\bibitem{TAPcomment} We note that this formulation is not suited to describe long-time glassy dynamics and effects such as aging.
\end{thebibliography}
\end{document}